# Glassy crystals with colossal multi-baroresponsivities

Kun Zhang[1,6,#], Zhe Zhang[1,2,#], Hailong Pan[3], Xueting Zhao[1,2], Ji Qi[1,2], Zhao Zhang[1,2], Ruiqi Song[1], Chenyang Yu[1,2], Biaohong Huang[1,2], Xujing Li[4,5], Huaican Chen[4,5], Changlong Tan[6], Wen Yin[4,5], Weijin Hu[1,2], Michael Wübbenhorst[3], Jiangshui Luo[3,7,*], Dehong Yu[8,*], Zhidong Zhang[1,2], and Bing Li[1,2,*]

[1]*Shenyang National Laboratory for Materials Science, Institute of Metal Research, Chinese Academy of Sciences, 72 Wenhua Road, Shenyang, Liaoning 110016, China.*

[2]*School of Materials Science and Engineering, University of Science and Technology of China, 72 Wenhua Road, Shenyang, Liaoning 110016, China.*

[3]*Laboratory for Soft Matter and Biophysics, Department of Physics and Astronomy, KU Leuven, Leuven, Belgium.*

[4]*Institute of High Energy Physics, Chinese Academy of Sciences, Beijing 100049, China.*

[5]*Spallation Neutron Source Science Center, Dongguan 523803, China.*

[6]*School of Materials Science and Chemical Engineering, Harbin University of Science and Technology, Harbin 150080, China.*

[7]*College of Materials Science and Engineering, Sichuan University, Chengdu 610064, China.*

[8]*Australian Nuclear Science and Technology Organisation, Lucas Heights, NSW 2234, Australia.*

[#]These authors equally contributed to the work.

[*] Corresponding authors: bingli@imr.ac.cn, jiangshui.luo@scu.edu.cn, or dyu@ansto.gov.au.

**As a nontrivial solid state of matter, the glassy-crystal state embraces physical features of both crystalline and amorphous solids, where a long-range ordered periodic structure formed by the mass centers of constituent molecules accommodates orientational glasses[1-3]. Here, we discover and validate a glassy-crystal state in 2-amino-2-methyl-1,3-propanediol (AMP, $C_4H_{11}NO_2$) by neutron scattering and complementary broadband dielectric spectroscopy (BDS) measurements. The freezing process of the dynamic orientational disorder is manifested at relaxation times well described by the Vogel-Fulcher-Tammann (VFT) law and the strongly frequency-dependent freezing temperature ranging from around 225 K at 0.1 Hz to above room temperature in the GHz region. At room temperature, the supercooled state is extremely sensitive to pressure such that a few MPa pressure can induce crystallization to the ordered crystal state, eventually leading to a temperature increase by 48 K within 20 s, a significant reduction of visible light transmittance from about 95% to a few percentages, and a remarkable decrease of electrical conductivity by three orders of magnitude. These ultrasensitive baroresponsivities might find their applications in low-grade waste heat recycling, pressure sensors and non-volatile memory devices. It is expected that glassy crystals serve as an emerging platform for exploiting exotic states of matter and the associated fantastic applications.**

Crystals and glasses are the most common solid states of matter and they have manifested themselves in distinct atomic structures[4,5]. A crystal has a well-defined atomic structure in which constituent units are arranged in a periodic pattern with certain crystallographic symmetries, whereas a glass (more specifically structural glass) usually lacks a positional and/or orientational order and resembles a frozen liquid. In contrast, a crystal is formed when cooling a melt below its solidification point while quenching a melt at a fast enough rate to avoid crystallization leads to supercooling and finally to a glassy state. Thus, crystals and glasses represent the ordered and disordered states of solids, respectively. As a hybrid solid state of matter, a glassy-crystal state combines both crystalline and glassy features in a single-phase material. A glassy-crystal state can be described with such a structure in which the mass centers of molecules develop a high-symmetry lattice and the molecular orientations are randomly frozen over the lattice sites, leading to static orientational disorder. Such a structure was firstly proposed in cyclohexanol by Adachi et al. about 60 years ago and a few potential systems have been reported later, but the exact microscopic picture hasn't been well established yet [1-3,6-8].

A glassy-crystal state can be obtained by quenching a plastic crystal just as quenching a melt to achieve a glass. In fact, a glassy crystal is usually formed at a moderate cooling rate. Plastic crystals are known as orientationally disordered molecular crystals in which molecules dynamically reorient over high-symmetry lattices[8-10]. In other words, the high symmetry of a plastic crystal originates from the dynamic nature of orientational disorder that is usually described by an isotropic reorientation mode. The typical timescale of such a disorder is around picosecond and thus the plastic crystal can be also regarded as an orientational liquid[11-13]. Upon crystallization, the alignment of the molecules leads to a lattice symmetry-breaking and ordered-crystal phases occur. This process is usually accompanied by huge latent heat due to the suppression of extensive molecular disorder, which enables their applications in conventional thermal energy storage[14,15]. Due to their intrinsic susceptibility to external pressure, the phase transition can also be triggered by pressure, which leads to colossal barocaloric effects that can provide a promising solution for next-generation solid-state refrigeration technology[11,12,16,17]. Between the glassy-crystal glass transition temperature and the plastic-to-ordered crystal transition temperature, the system is metastable and resides in a supercooled plastic crystal state[2]. Resembling the well-known volume versus temperature curve for a glass-forming liquid[4], we schematically plot the counterpart of the glassy-crystal glassy transition in **Fig. 1a**. As far as technological applications are concerned, the thermodynamic metastable nature of the glassy and/or supercooled state implies ultra-sensitivity to external stimuli like pressure[18].

Here, we reveal and validate the existence of a glassy-crystal state in AMP through the investigations of both crystal structures and molecular dynamics by combining neutron scattering and complementary BDS measurements. Furthermore, we exploit the ultra-barosensitive thermal, optical, and electrical properties of the AMP in association with this well-defined glassy-crystal state. It is anticipated that these unique properties may be generic to such a nontrivial state of matter.

AMP has been widely applied as a chemical agent in many industrial processes[19]. It is known that the thermodynamic equilibrium state of AMP at room temperature is a crystal with a monoclinic structure that undergoes a phase transition to a cubic structure upon heating at about 353 K[10,20,21]. This phase transition is regarded as a plastic-crystal transition accompanied by huge latent heat, ~ 245 J g$^{-1}$. In contrast, during the cooling process, the plastic-crystal state is maintained down to room temperature, previously considered as a supercooled state[21, 22]. **Fig. 1b** shows the heat flow data over the temperature region between 100 and 425 K for the cooling and heating processes monitored using a differential scanning calorimetry (DSC). Here, we identify four solid states, i.e., plastic crystal (I), ordered crystal (II), supercooled plastic crystal ($I_{sc}$), and glassy crystal ($I_g$), which are labeled in **Fig. 1b**, respectively. During the cooling process after solidification at about 382 K, the I state enters into the regime of the $I_{sc}$ state and further into the $I_g$ state at 225 K. Upon heating, the $I_g$ state transforms to the $I_{sc}$ state at 225 K, followed by crystallization to the state II at 280 K with entropy changes of −581.6 J kg$^{-1}$ K$^{-1}$. The II state undergoes the plastic-crystal transition to the I phase at 361 K with entropy changes of 661.8 J kg$^{-1}$ K$^{-1}$. The natures of these phases will be clarified in detail below.

The evolution of crystal structures of these phases is monitored using temperature-dependent neutron powder diffraction (NPD) and X-ray diffraction (XRD). As shown in **Fig. 1c**, the NPD pattern of the partially deuterated sample taken at 300 K is identified as the monoclinic phase ($P2_1/n$, $a = 8.6194$ Å, $b = 11.0419$ Å, $c = 6.1207$ Å, and $\beta = 93.6°$). Upon heating to 363 K, only very few Bragg peaks are observed, which indicates a high-symmetry crystal structure indexed by a body-centered cubic (BCC) structure ($Im3m$, $a = 6.8194$ Å). After cooling back to room temperature, the BCC phase is maintained with a reduced lattice constant. This BCC phase persist upon further cooling down to 200 K, which disproves the existence of a structural glassy state. These observations are further confirmed by the temperature-dependent lab XRD patterns and the diffraction pattern collected at Pelican (**Extended Data Figs. 1a and 1b**). Furthermore, the unit cell volume of the BCC and monoclinic phases are determined at room temperature (**Extended Data Figs. 1c and 1d**). It is suggested that the supercooled BCC phase has a 9.1% larger unit cell volume than the monoclinic phase, which indicates a "free volume" of this glassy-crystal state, similar to structural glasses[4]. As Raman scattering reflects the lattice symmetry as well, temperature-dependent Raman spectra collected under similar heating and cooling conditions are plotted in **Extended Data Fig. 2** from room temperature to 360 K in a step of 10 K. In terms of previous reports, the pattern at 300 K suggests the ordered crystal phase with the monoclinic structure[23]. Consequently, the Raman spectra underpin the XRD and NPD results.

Then, one wonders how molecules distribute over such a BCC lattice without inducing a lattice symmetry-breaking. To this end, we investigate the phonons and reorientation dynamics using inelastic neutron scattering (INS) and quasi-elastic neutron scattering (QENS), respectively. These dynamics measurements were performed on a hydrated sample at the time-of-flight cold neutron spectrometer Pelican (**Methods**). In **Fig. 1d**, the phonon density of state (DOS) is plotted up to energy transfer ($E$) of 70 meV for the I, II, and $I_{sc}$ states. At room temperature, the sample crystallizes in the monoclinic phase, whose spectrum is typical of crystalline materials with a steep low-energy part and some well-defined peaks imposed over the broad pattern in the high-energy part. At 363 K, the compound becomes the plastic crystal with a broad and featureless pattern, just as observed for other classical plastic crystals like neopentylglycol[11], $NH_4I$[17], and $NH_4SCN$[24]. With cooling down to room temperature, the supercooled phase displays a slightly sharpened pattern. The inset highlights the low-energy parts for these three states as a function of $E^2$. The slopes of the plots are inversely proportional to the phonon velocities to the power of 3 and thus the bulk modulus according to the Debye model. It is plausible that the ordered-crystal phase and the plastic-crystal phase have the largest and the smallest bulk moduli while the supercooled state falls into the middle.

QENS is a desirable tool to study molecular reorientation dynamics of hydrogen-containing systems due to the huge incoherent scattering cross-section of hydrogen atoms[25]. QENS signal usually appears as a broad spectrum imposed underneath the elastic line and is described by a Lorentzian function[11,17,24]. Shown in **Figs. 2a**, **2b**, and **2c** are the dynamic structure factors $S(Q, E)$ as a function of momentum transfer ($Q$) and energy transfer ($E$) with incident neutron energy $E_i$ = 3.7 meV for these three states, respectively. The intense strips centered at $E = 0$ represent the elastic line, which contains most of the scattering intensity whereas a less-intense signal spreading out from $E = 0$ is the QENS intensity. At 293 K, the compound crystallizes in the monoclinic phase, whose $S(Q, E)$ is described by the elastic lines and suggests the nature of an ordered-crystal state. At 363 K, the spectrum is typical of a plastic-crystal state, where a strong QENS signal is observed. In contrast, the supercooled state shows a very weak but visible QENS signal. These differences become clearer in the constant $Q$ spectra at $Q = 1.6$ Å$^{-1}$ as shown in **Figs. 2d, 2e**, and **2f**. The spectrum of the ordered crystal is well fitted to a combination of a delta function and a constant background, convoluted with the instrumental resolution. The absence of a Lorentzian function in this fitting indicates the lack of reorientation motion in the present frequency window, in agreement with the ordered-crystal structure. In the plastic-crystal phase, an intense Lorentzian function is required to reproduce the data and represents active reorientation motions. The experimental elastic incoherent scattering factor (EISF) is obtained and fitted to an isotropic reorientation model with the rational radius of 0.68(15) Å (**Extended Data Fig. 3a**). The full width at half maximum (FWHM) of the Lorentzian function for the plastic-crystal phase is ~ 0.6 meV (**Extended Data Fig. 3b**). For the supercooled state, one order of magnitude weaker QENS suggests that most of the reorientation motions are suppressed. The distinct spectra of the plastic-crystal state and the supercooled state are remarkably impressive since they have identical lattice symmetry. Our QENS data unambiguously verify the frozen nature of the supercooled state even if with the same crystal symmetry as the plastic-crystal phase.

The above QENS data are collected with an energy resolution of about 130 μeV, which corresponds to a frequency domain of about 30 GHz. In this time window, we observe that the molecular reorientation motions are frozen in the supercooled phase. To study the dynamic freezing process on a wider time scale, we apply BDS to investigate the frequency responses covering a range of several orders of magnitude of frequency[8]. **Figs. 3a** depicts the surface plots of the real and the imaginary part of the complex dielectric permittivity (dielectric loss) $\varepsilon''$ ($f$, $T$) measured upon cooling at frequencies from $10^{-1} - 10^6$ Hz, which reveals a frequency-independent drop of $\varepsilon''$ at 382 K corresponding to the transition from the liquid to plastic crystalline state in agreement with the heat flow data (**Fig. 1b**). Further cooling yields a monotonous decrease of the dielectric loss along with a continuous shift of the dominant α-relaxation process down to the glassy crystal state. This scenario is also seen in the temperature dependence of two other quantities, the dielectric permittivity $\varepsilon'$ and the real part of the complex electrical conductivity ($\sigma'$) at $f$ = 0.10 Hz, all revealing a high-temperature first order event at 382 K and a dynamic glass transition at the low temperature end (**Fig. 3b**). To follow the dynamic glass transition, manifested by the α-relaxation peak in **Fig. 3a** to shorter timescales, additional RF dielectric spectra were taken to close the gap between "standard" dielectric spectroscopy and the GHz time scale of the QENS data. **Fig. 3c** displays dielectric loss spectra covering the extended frequency range from $10^{-1}$ to $10^9$ Hz at selected temperatures from 393 down to 193 K. Visual inspection of the curves confirms the expected continuous nature of the dynamic freezing in the entire frequency range, an observation that is qualitatively confirmed by the smooth temperature dependence of the relaxation time $\tau_\alpha$ upon cooling presented in **Fig. 3d** (blue symbols). Moreover, all relaxation time data obey very well the VFT law, $\tau(T) = \tau_\infty exp(E_V/R(T - T_V))$, where $E_V$, $R$, and $T_V$ denote the "Vogel activation energy", the gas constant, and the Vogel temperature, respectively. The surface plots of other quantities ($\varepsilon'$ and $\sigma'$) upon cooling are shown in **Extended Data Figs. 4a and 4b.** And the temperature-dependent dielectric properties at other selected frequencies are represented in **Extended Data Figs. 4c and 4d**.

Based on our extensive structural and dynamic data, we can confirm the existence of supercooled plastic crystal and glassy-crystal states for AMP. At room temperature, the as-cooled BCC phase is a supercooled plastic crystal state where faster reorientation motions than $10^9$ Hz are frozen while slower ones are still active. The freezing does not induce a lattice symmetry-breaking. Usually, one defines the glassy transition temperature by the freezing temperature under a slow enough process. Here, $T_g$ (about 225 K) is determined from the heat flow measurement. Glassy crystal is a long-sought nontrivial state of matter[1-3,6-8], but the exact microscopic image has not been well clarified. The systematic structural and dynamics investigation of AMP has shed more light on this topic which offers an emerging playground for glasses physics.

Now we explore the response to external pressure of the glassy crystal given that pressure is a universal governing factor of an atomic system. The in-situ pressure responses of AMP are studied using DSC, XRD, and Raman scattering. Shown in **Fig. 4a** are the heat flow data during a complete heating-cooling-pressurization process. Initially, the compound is heated from 273 to 393 K under 0.1 MPa and two endothermic peaks appear, corresponding to the ordered crystal to plastic crystal phase transition and the melting, respectively. Subsequently, the compound is cooled down to 273 K and there is only the phase transition from liquid to plastic crystal. At 273 K, the supercooled state is characteristic of an exothermic peak under a pressure of 6.7 MPa. The latent heat related to this peak is estimated to be 134 J g$^{-1}$. This value is close to the latent heat during the temperature-induced transition from $I_{sc}$ to II phases (**Fig. 1b**), suggesting the occurrence of the pressure-induced phase transition. In **Extended Data Fig. 5**, this temperature-pressure cycle has been repeated with varying conditions, indicative of excellent reproducibility. Shown in **Fig. 4b** are the pressure-dependent in-situ Raman spectra. Under ambient pressure, only one strong peak is found at 765 cm$^{-1}$. Upon applying a pressure, five strong peaks are observed at 529, 560, 779, 1052, and 1176 cm$^{-1}$, which correspond to the C-C-N bending, O-C-C bending, C-N stretching, C-O vibration, and C-N stretching modes[23], respectively. Along with the temperature-dependent Raman spectra (**Extended Data Fig. 2**), it is confirmed that crystallization takes place. Moreover, the in-situ XRD also confirms the phase transition with pressure (**Extended Data Fig. 6**).

Harnessing the pressure-induced crystallization from the supercooled plastic-crystal state, we demonstrate the thermal, optical as well electrical applications. To make use of the huge latent of the crystallization, a concept of barocaloric thermal battery is proposed. It aims to tackle the thermal energy paradox that the production of heat accounts for more than 50% of global final energy consumption[26] while the waste heat potential analysis reveals that 72% of the global primary energy consumption is lost after conversion mainly in the form of low-grade waste heat[27]. The low-grade waste heat is harvested and stored in the thermal battery and then reused on demand upon pressurization. The barocaloric thermal battery includes three stages as shaded in **Fig. 4a**, i.e, thermal charging upon contacting the waste heat source, heat storage in the supercooled state, and the pressure-induced exothermic process. The last step is directly demonstrated on a sample encapsulated with a thermocouple using ethylene-vinyl acetate copolymer tapes. The real-time temperature change is recorded as a function of time. As can be seen in **Fig. 4c**, a huge temperature rise of about 48 K is achieved within 20 s. This process has been also visualized by an infrared camera as shown in the inset (**Supplementary Video S1**). In air, an as-prepared supercooled sample on a quartz glass plate is captured in dark blue at 297 K. When the sample is needled, a few bright spots appear immediately indicating a temperature increase. After a few seconds, the high-temperature zones emerge, and a maximum temperature of ~ 333.8 K is reached at 20 s before the temperature decays to the original state in about 60 s. Such performances are ranked the best among all current active thermal energy storage materials, as listed in **Extended Data Table 1**[24,28-31].

In addition to the thermal battery, the glassy crystal can also serve as a pressure-to-heat sensor and the translated heat signal is well detected in the form of IR radiation. Indeed, pressure-to-electrical and pressure-to-optical sensors are feasible, too. An as-prepared supercooled state is transparent to visible light while the ordered crystal is white (**Extended Data Fig. 7** and **Supplementary Video S2**). To quantitatively characterize the optical property, the ex-situ transmittance to visible light is illustrated in **Fig. 4d** in the wavelength range from 200 to 800 nm. The transmittance of the supercooled plastic crystal shows a sharp increase at about 250 nm, with almost 100% transmittance for the whole wavelength range above 250 nm. By contrast, the ordered crystal only transmits below 10% in the whole spectrum range. For example, the transmittance at 300 nm is lowered from ~ 95.2% to ~ 1.5% upon the pressure-driven phase transition from the supercooled state to the ordered state. This ultra-optical sensitivity may serve as an excellent optical filter. Similar to optical transmittance, the electrical conductivity displays a $10^3$ drop at about 300 K from the supercooled state to the ordered crystal state (see the inset in **Fig. 4d**). The thermal, optical, and electrical ultrasensitive responses to pressure make AMP an ideal material for fabricating high-performance force-to-heat, force-to-optical, and force-to-electrical sensors. The driven pressure of few MPa is many orders of magnitude lower than leading sensor materials like $VO_2$[32].

The third promising application is the new-concept non-volatile memory device based on the concurrent thermal/pressure responsibility. Following the thermal-pressure cycle shown in **Fig. 1a**, the ordered-crystal state can be defined as the logic state 0, which can be written using thermal annealing to the supercooled plastic-crystal state as the logic state 1, which is rewritten to be the logic state 0 by pressure. These two states survive after the removal of pressure and are well electrically and/or optically read. In addition, by dedicate controlling the magnitude and duration of the stress, it is possible to achieve various conducting states between that of supercooled plastic crystal and the ordered crystal, leading to a new type of multi-level memories or even synapse devices used for neuromorphic computation.

Towards real applications, reliability is the one of most critical issues, which involves reproducibility and stability. The fabrication of the supercooled plastic crystal state has been successfully tested six times in the temperature region from 273 to 393 K, as shown in **Extended Data Fig. 8a**. In addition, the sample dependence is also considered. The heating-cooling-storage process has been readily repeated in three different samples (**Extended Data Figs. 8b, 8c,** and **8d**). As sensitivity and stability are normally a pair of contradictory performances, the long-term aging stability of the prepared supercooled plastic crystal state of AMP samples is monitored by ex-situ XRD, in-situ Raman scattering, and camera. As shown in **Extended Data Fig. 9**, the supercooled state can survive up to a half year in a glove box.

In summary, the crystal structures, atomic dynamics, and phase transitions of AMP have been systematically investigated. The existence of a nontrivial glassy-crystal state has been verified in a wide temporal domain from $10^{-1}$ to $10^6$ Hz by BDS measurements and further to $10^{10}$ Hz by QENS. The pressure-induced crystallization of the supercooled plastic crystal state enables ultrahigh-performance barocaloric thermal battery for the utilization of low-grade waste heat, ultrasensitive pressure sensors, and non-volatile memory devices. These baroresponsivities are most likely universal for glassy crystals. Our work is expected to pave an emerging route to explore glass physics and novel applications in association with the nontrivial glassy crystals.

**Methods**

*Material:* AMP (2-amino-2-methyl-1,3-propanediol; $H_2N(CH_3)C(CH_2OH)_2$; purity: ≥99%; CAS number: 115-69-5) samples were purchased from Sigma-Aldrich. The partially deuterated sample was prepared by replacing hydrogen atoms associated with N and O sites using $D_2O$ (99.9 atom % D, Aldrich) *via* a secondary deuteration. To a flask with 3.6 mL $D_2O$ (99.9 atom % D, Aldrich), 4.00 g of the sample was added. It was then sealed and stirred at 313 K for 48 h. After that, it was heated at 383 K for 2 h while purged with pure nitrogen gas (99.999%) to remove the solvent $D_2O$. Then, the sample, together with the flask, was frozen at around 254 K in a refrigerator for 2 h before it was purified for 1 h on a Schlenk vacuum line at room temperature to remove the residual $D_2O$. The fine powders were again dried on a Schlenk vacuum line at room temperature for 1 h. The as-obtained product was deuterated again to improve the purity. The final deuterated ratio of ~28.6% was determined by $^1$H-NMR (Advance II-400 MHz, Bruker; solvent: DMSO-*d*6).

***Pressure-induced temperature changes*:** We directly measured the temperature change of the sample during a pressure-induced phase transition using thermography and thermocouple, respectively. The pressure was applied to a supercooled sample using a needle. For a supercooled sample exposed in air, the temperature evolution of the sample during the pressure-induced phase transition was recorded using an infrared camera (Fluke TiX580). For a supercooled sample encapsulated by ethylene-vinyl acetate copolymer tapes, the pressure-induced temperature changes at room temperature were measured with a K-typed thermocouple, which was inserted into the center of the sample.

***Calorimetric characterizations*:** Heat flow data under ambient pressure of AMP were collected using two differential scanning calorimeters, μDSC7, Setaram for scans at 2 K min$^{-1}$) as well as NETZSCH DSC 200F3 for scans at 10 K min$^{-1}$. The phase-transition temperature was defined as the temperature at which the heat flow peaked. The pressure-dependent heat flow data were collected using a high-pressure differential scanning calorimeter (μDSC7, Setaram). The sample was enclosed in a high-pressure vessel made of Hastelloy. Firstly, the sample of the supercooled state was obtained using the μDSC7 under atmospheric pressure and was kept at 273 K. For the constant temperature process, the samples of AMP were pressurized from 0.1 to 8.7 MPa. Similar procedures have been applied to $NH_4SCN$[24] and $NH_4I$[17].

***XRD:*** The ex-situ XRD data were collected on the II and $I_{sc}$ states at room temperature at an X-ray diffractometer (Miniflex, Rigaku) with $K_{α1}$ radiation of Cu. The in-situ XRD experiments were performed at another X-ray diffractometer (Bruker D8) equipped with an HPC 900 chamber with $K_{α1}$ radiation of Cu for temperature-dependent measurements and with $K_{α1}$ radiation of Mo for pressure-dependent measurements. For the temperature-dependent experiments, the sample was heated from 303 to 363 K and then cooled back to 303 K. For the pressure-dependent experiments, an as-prepared supercooled sample was measured under ambient pressure and nitrogen gas pressure of 8 MPa at 303 K, respectively. The ex-situ XRD data were analyzed using a Le Bail refinement method[33].

*NPD:* NPD experiments were performed at the Multi-Physics Instrument (MPI) of China Spallation Neutron Source (CSNS) in China[34]. A partially deuterated sample amount of 0.65 grams was put into a vanadium can ($\phi$ = 9 mm, $L$ = 75 mm). The sample was heated from 300 to 363 K, then cooled to 200 K. Constant-temperature scans were carried out at 300, 363, 300 (cooled back), and 200 K, respectively. At each temperature, the counting persisted for 3 h.

*Raman Scattering:* Raman spectra were collected using a commercial Raman system (Horiba Labram HR Evolution) under the normal incidence of a helium-neon laser ($\lambda$ = 532 nm). The laser beam was focused on the samples by a × 50 objective (numerical aperture 0.6). The beam diameter was about 1μm. The sample was located in a continuous-flow liquid-nitrogen cryostat. A calibrated Linkam heating-cooling stage was utilized to control sample temperature via a thermocouple attached to the sample holder. High pressure was accessed through a membrane diamond anvil cell equipped with beryllium copper gaskets with holes of about 500 μm serving as the sample chamber. Refer to the previous results for detail[35].

*INS/QENS:* The INS and QENS experiments of hydrated AMP were conducted at the time-of-flight neutron spectrometer Pelican of Australian Center for Neutron Scattering (ACNS) of Australian Nuclear Science and Technology Organization (ANSTO) in Australia[36]. The instrument was configured with an incident neutron wavelength of 4.69 Å, affording an incident energy of 3.72 meV with an energy resolution of 0.135 meV at the elastic line. The AMP powder sample was put into a quartz tube having 4 mm inside diameter and 0.5 mm wall thickness. The quartz tube is inserted into a sample can made of a niobium sheet. The sample handling process was carried out in a glove box filled with dry nitrogen gas under atmosphere pressure. The measurements were carried out at 293, 363, and 293 K during the heating-cooling process. The empty can was measured in the same conditions for background subtraction. In addition, a standard vanadium sample was also measured for detector normalization and to determine the energy resolution function. The data reduction, including background subtraction and detector normalization, was performed using the Large Array Manipulation Program (LAMP)[37], while the sliced QENS spectra were analyzed with the Pan module built-in the Data Analysis and Visualization Environment (DAVE)[38].

***QENS analysis and reorientation dynamics***: Avoiding Bragg peaks, one-dimensional $Q$-sliced data were extracted. The fitting functions included a linear background function, a Lorentzian function (for supercooled plastic crystal state and plastic crystal state), and a Delta function, which were convoluted with the instrumental resolution. These components represented background, quasi-elastic scattering intensity ($I_{QENS}$), and incoherent elastic scattering intensity ($I_{elastic}$), respectively. Elastic Incoherent Structure Factor (EISF) is defined as the ratio of elastic scattering to the total scattered intensity[34],

$$\text{EISF} = \frac{I_{elastic}}{I_{elastic} + I_{QENS}} \quad (1)$$

A scaling factor $s$ was introduced to describe the multiple scattering[39]. An isotropic rotational model (Equation 2)[40,41] was used to describe the H atoms dynamics of the plastic crystal phase at 363 K, read as

$$\text{EISF}_{isotropic} = s * \left[\frac{\sin(Qr)}{Qr}\right]^2 \quad (2)$$

Here, $r$ is the rotational radius, which was estimated to be 0.68(15) Å.

***Dielectric relaxation spectroscopy:*** Dielectric properties ($\varepsilon'$, $\varepsilon''$ and $\sigma'$) were measured using a Broadband Dielectric Spectrometer (Novocontrol GmbH) with an active sample cell based on a high-resolution impedance analyzer (Novocontrol Alpha) at frequencies between $10^{-1}$ and $10^6$ Hz. The sample was placed in a stainless-steel liquid cell (diameter: 20.00 mm; electrode spacing: 1.560 mm) realizing a parallel plate geometry. For the upper frequency range from $10^6$ to $10^9$ Hz, a Hewlett Packard HP 4291B network analyzer in combination with a calibrated coaxial line and a Novocontrol RF-cell (parallel plates of 5 mm in diameter with 100 μm electrode separation) was used. The sample temperature was controlled by means of a Novocontrol Quatro cryosystem allowing temperature control with an accuracy of ± 0.01 K *via* a dry nitrogen gas flow derived from liquid nitrogen. To ensure complete filling of the cell, a fresh sample was first heated from room temperature to 393 K, followed by stabilization at this temperature for 15 min to eliminate any possible thermal history. The sample was then cooled down to 193.15 K at a cooling rate of 10 K min$^{-1}$. Afterwards, the sample was heated again to 393 K and then once again cooled down to 193 K with a temperature interval of 2 K (heating/cooling rate: 10 K min$^{-1}$) for the dielectric spectroscopy measurements. More details can be found in Refs. 42, 43, and 44.

***Optical measurement:*** Optical transmission spectrums with light wavelengths ranging from 200 to 800 nm were performed on a UV-VIS spectrophotometer (Hitachi U-3010). A double-side polished quartz substrate was used as the reference for the measurement.

**Acknowledgments**

The work conducted in the Institute of Metal Research was supported by the Key Research Program of Frontier Sciences of Chinese Academy of Sciences (Grant no. ZDBS-LY-JSC002), CSNS Consortium on High-performance Materials of Chinese Academy of Sciences, the Young Innovation Talent Program of Shenyang (Grant no. RC210435), the Ministry of Science and Technology of China (Grant nos. 2021YFB3501201 and 2020YFA0406002), the National Natural Science Foundation of China (Grant nos. 11934007, 52001101 and 61974147), and the International Partner Program of Chinese Academy of Sciences (Grant no. 174321KYSB20200008). Jiangshui Luo and Michael Wübbenhorst thank the funding from the National Natural Science Foundation of China (Grant no. 21776120) and the Research Foundation-Flanders (FWO) (Grant no. G0B3218N). We thank Sucheng Wang and Changji Li for their help with the in-situ XRD experiment, Shuai Huang for the thermal infrared image measurements, as well as Bo Huang for the deuteration of the sample. We thank Prof. Jie Pan and Prof. Si Lan for their valuable discussion.

**Author Contributions**

B.L. proposed the project. K.Z., Zhe Zhang, J.Q., Zhao Zhang, R.S., C.Y. and C.T. carried out the in-house structural, thermal and Raman scattering experiments. Zhe Zhang was co-supervised by B.L. and Zhidong Zhang. X.Z., X.L., H.C. and W.Y. performed neutron diffraction measurements. D.Y., Zhe Zhang and B.L. conducted INS/QNES measurements. J.Luo and M.W. designed the BDS experiments. H.P., M.W. and J. Luo performed BDS measurements and analyzed the BDS data. J.Luo and M.W. wrote the part related to BDS and revised the manuscript. J. Luo supervised and analyzed the deuteration of the sample. B.H. and W.H. measured the optical transmittance. Zhao Zhang and K.Z. analyzed the diffraction data. Zhe Zhang and K.Z. analyzed the QENS data. K.Z. and B.L. wrote the manuscript with input from all authors.

**Data availability**

The data of the present study are available from the corresponding authors upon reasonable request.

**Competing interests**

The authors declare no competing interests.

**Additional information**

Supplementary information is available for this paper at XXX.

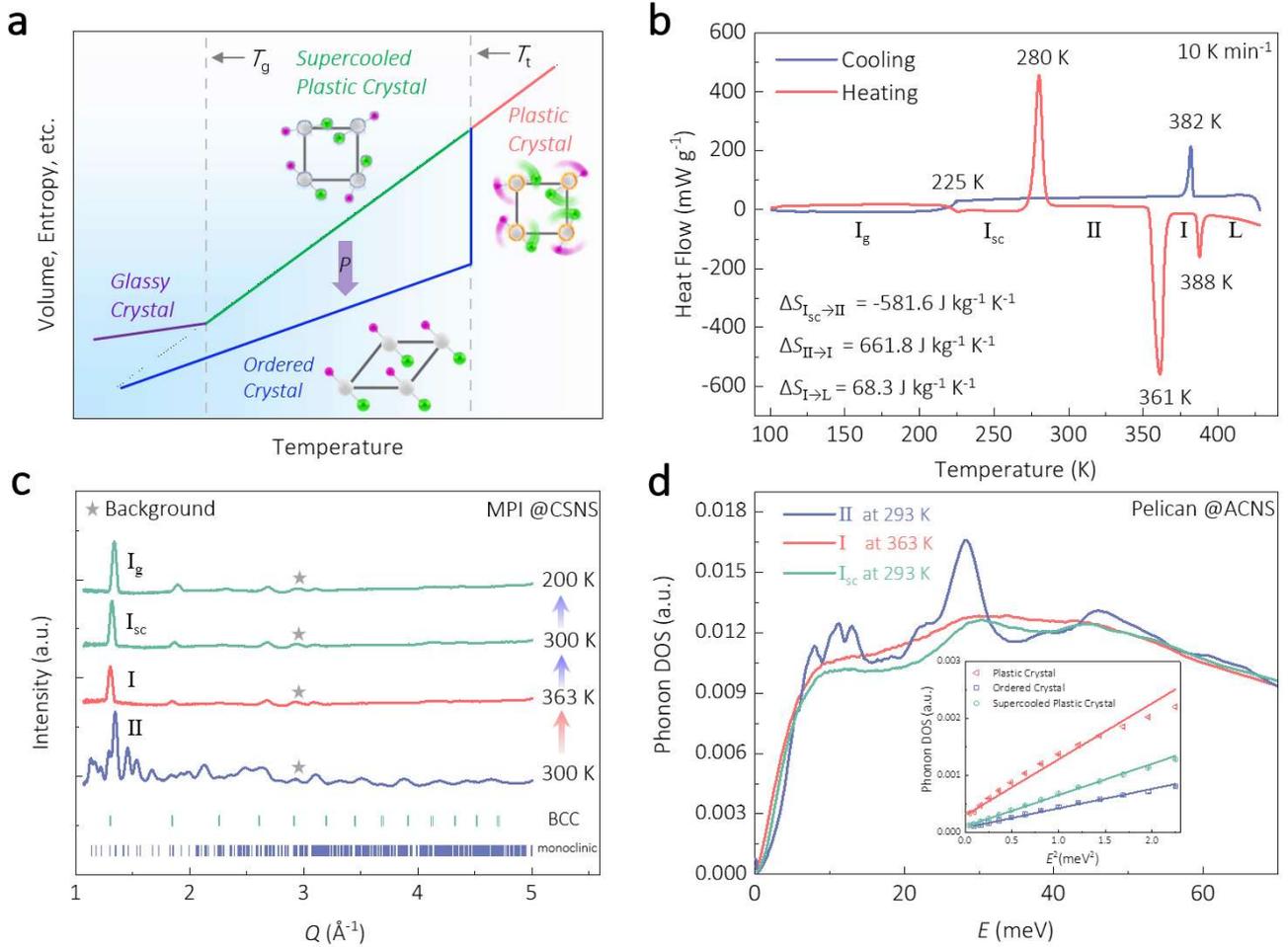

**Fig. 1 Phase transitions of AMP under ambient pressure. a**. A schematic diagram for a glassy-crystal glassy transition. The ordered-crystal state is heated to the plastic-crystal state at $T_t$. At cooling down, the supercooled plastic-crystal state emerges and finally the glassy-crystal state is formed at $T_g$. The arrow with $P$ means pressure induces crystallization of supercooled plastic crystal. Schematic atomic structures of these states are also illustrated. **b**. Heat flow data on cooling and heating of AMP under ambient pressure over the temperature region between 100 and 425 K. **c**. NPD patterns at different temperatures on a partially deuterated sample of AMP obtained at MPI of CSNS. The red (blue) arrows represent the heating (cooling) runs. Positions of Bragg peaks of two phases are ticked. **d.** Phonon DOS of AMP at different temperatures obtained at Pelican of ACNS. The inset highlights the low-energy parts as a function of $E^2$.

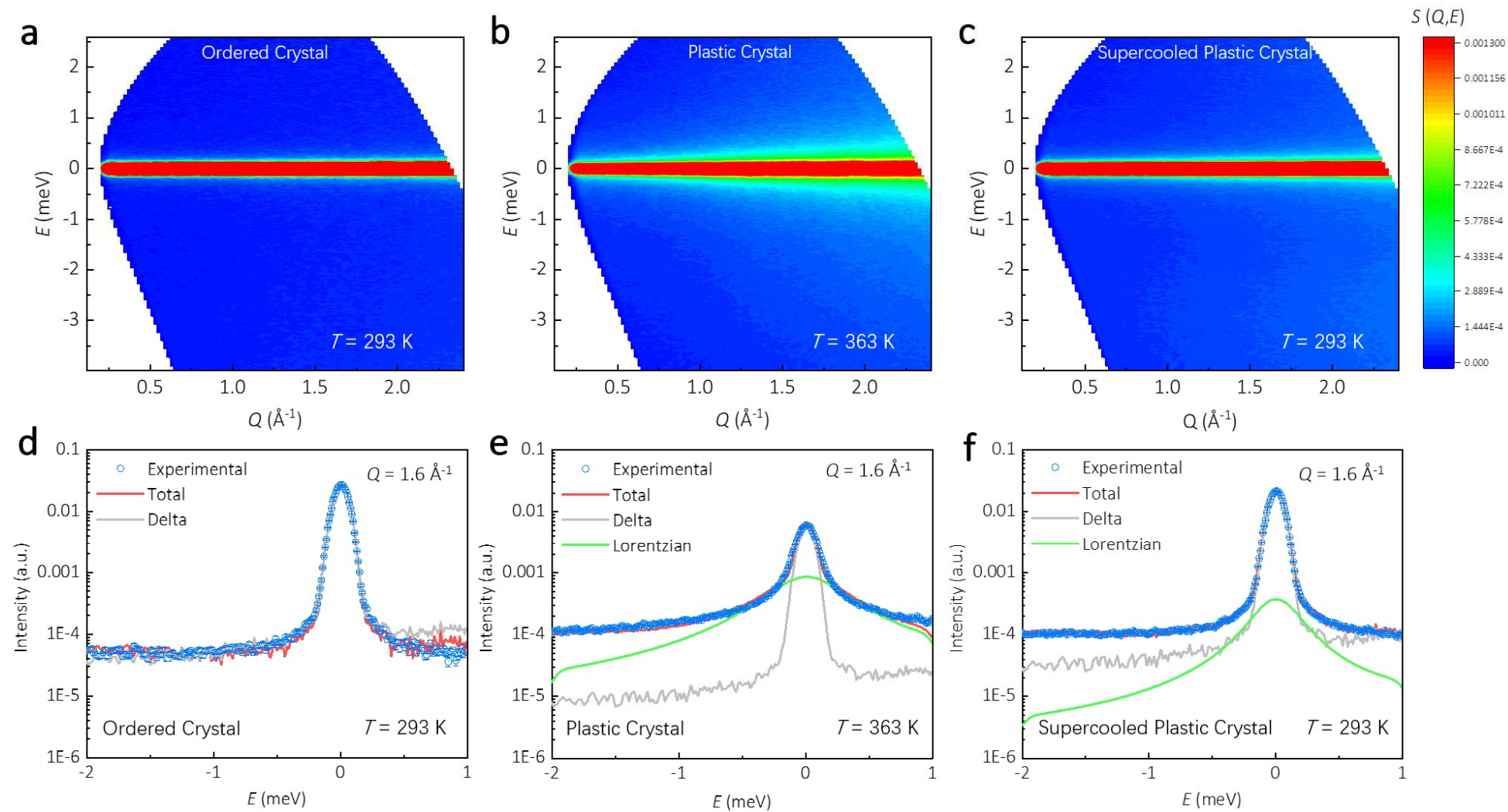

**Fig. 2 Reorientation dynamics of AMP.** The dynamic structure factor $S(Q, E)$ at (**a**) 293, (**b**) 363, and (**c**) 293 K (cooled back) at Pelican. Spectral fitting of the sliced $S(Q, E)$ for the $Q$ range of [1.55, 1.65] Å$^{-1}$ at (**d**) 293, (**e**) 363, and (**f**) 293 K (cooled back), respectively.

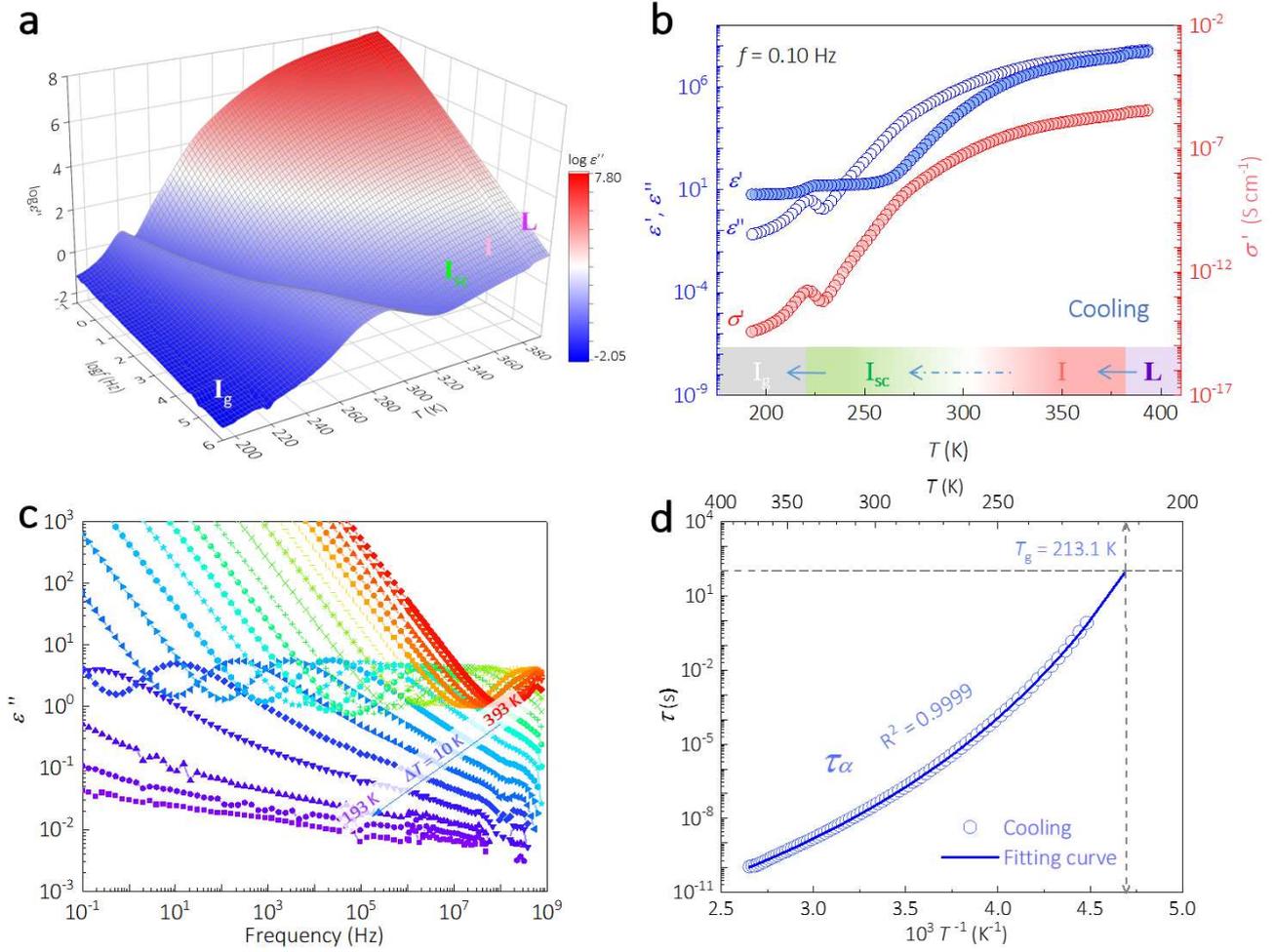

**Fig. 3 Broadband dielectric spectra across phase transitions. a.** Surface plot of the dielectric loss $\varepsilon''$ ($f$, $T$) (imaginary part of the complex dielectric permittivity) upon cooling at frequencies from $10^{-1} - 10^6$ Hz. **b**. The isochronal plot of $\varepsilon'$, $\varepsilon''$ as well as the real part of the complex electrical conductivity ($\sigma'$) at $f = 0.10$ Hz at cooling. **c**. Dielectric loss spectra in the extended frequency range from $10^{-1}$ to $10^9$ Hz at selected temperatures. **d**. Arrhenius diagram showing the relaxation time of the dominant α-relaxation process for the cooling process (blue symbols), fitted to the VFT law. The extrapolated temperature at $\tau = 100$ s defines the glassy-crystal glass transition temperature $T_g$ from the dielectric properties, which is 213.1 K, slightly different from 225 K defined in heat flow data.

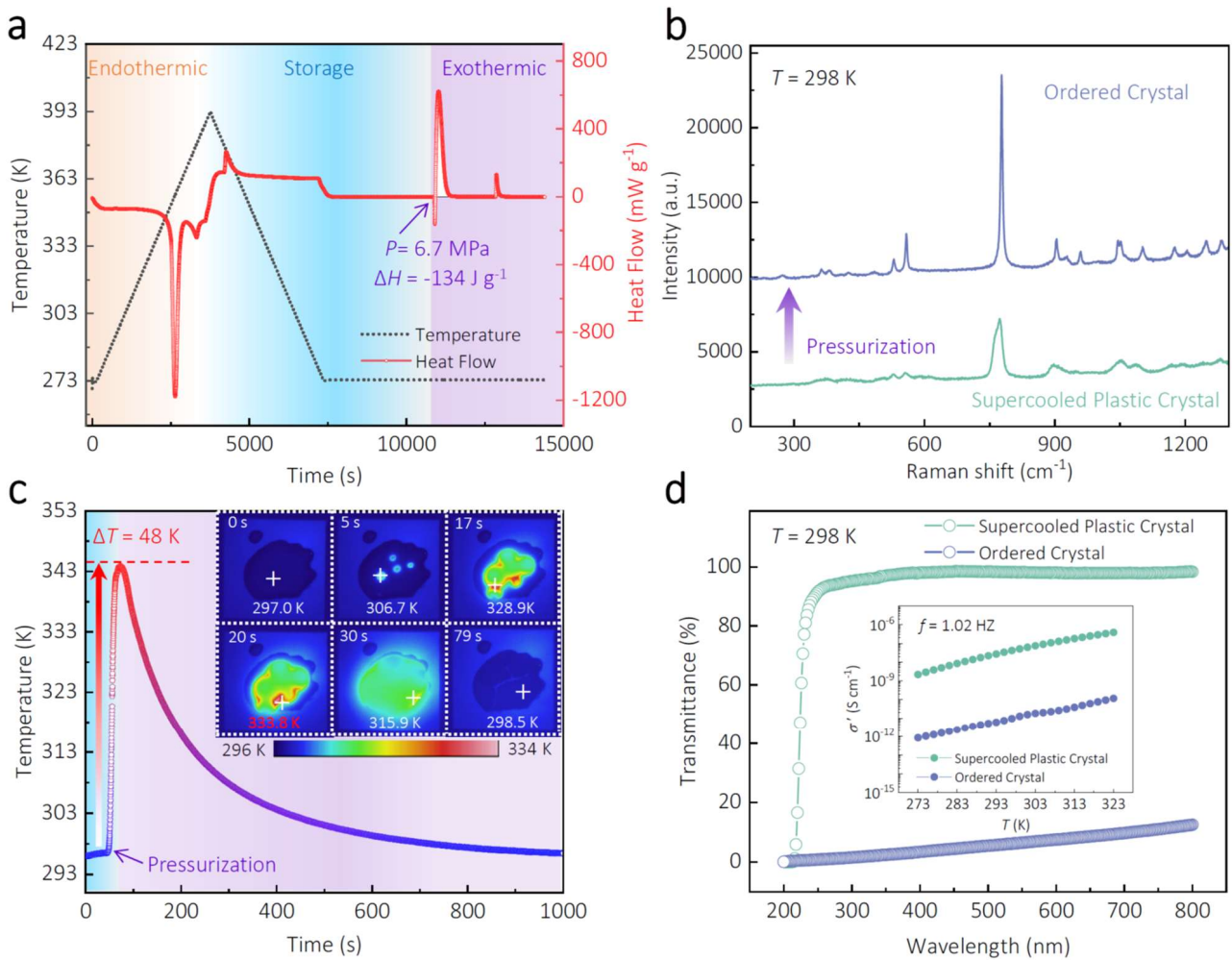

**Fig. 4 Pressure-induced crystallization of the supercooled state and its thermal, optical and electrical responses. a.** Heat flow variations as a function of time for the thermal battery cycle, including the endothermic, storage and pressure-induced exothermic processes. The moment for applying pressure is arrowed. **b.** In-situ Raman spectra at room temperature. **c**. Direct measurements of the temperature rise at the pressure-induced crystallization of the supercooled plastic crystal. The inset highlights the temporal evolution of the temperature monitored by an infrared camera upon applying pressure to the supercooled plastic crystal state. The bright spots appearing at 5 s represent the immediate changes in the needled zones. **d**. Ex-situ optical transmittances in the wavelength from 200 to 800 nm for the supercooled state and the ordered crystal. The inset shows the real part of the complex electrical conductivity $\sigma'$.

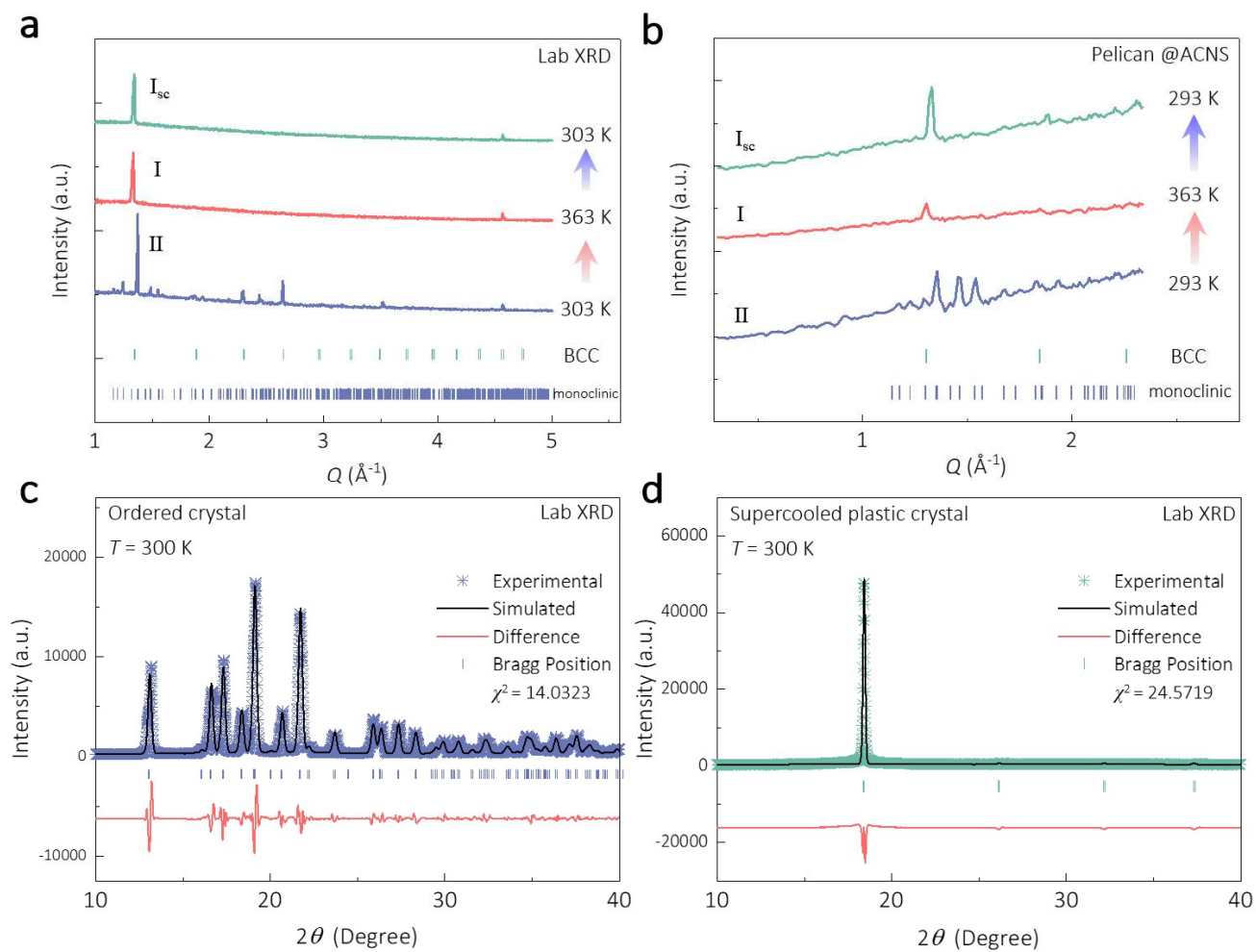

**Extended Data Fig. 1 Additional diffraction data. a.** In-situ XRD patterns of AMP at different temperatures. Constant-temperature scans were done at 303, 363, and 303 K (cooled back). **b**. Elastic components of neutron scattering at the Pelican of ACNS. **c** and **d**, Lab ex-situ XRD patterns and refinements at room temperature of the ordered-crystal state and the supercooled plastic-crystal state.

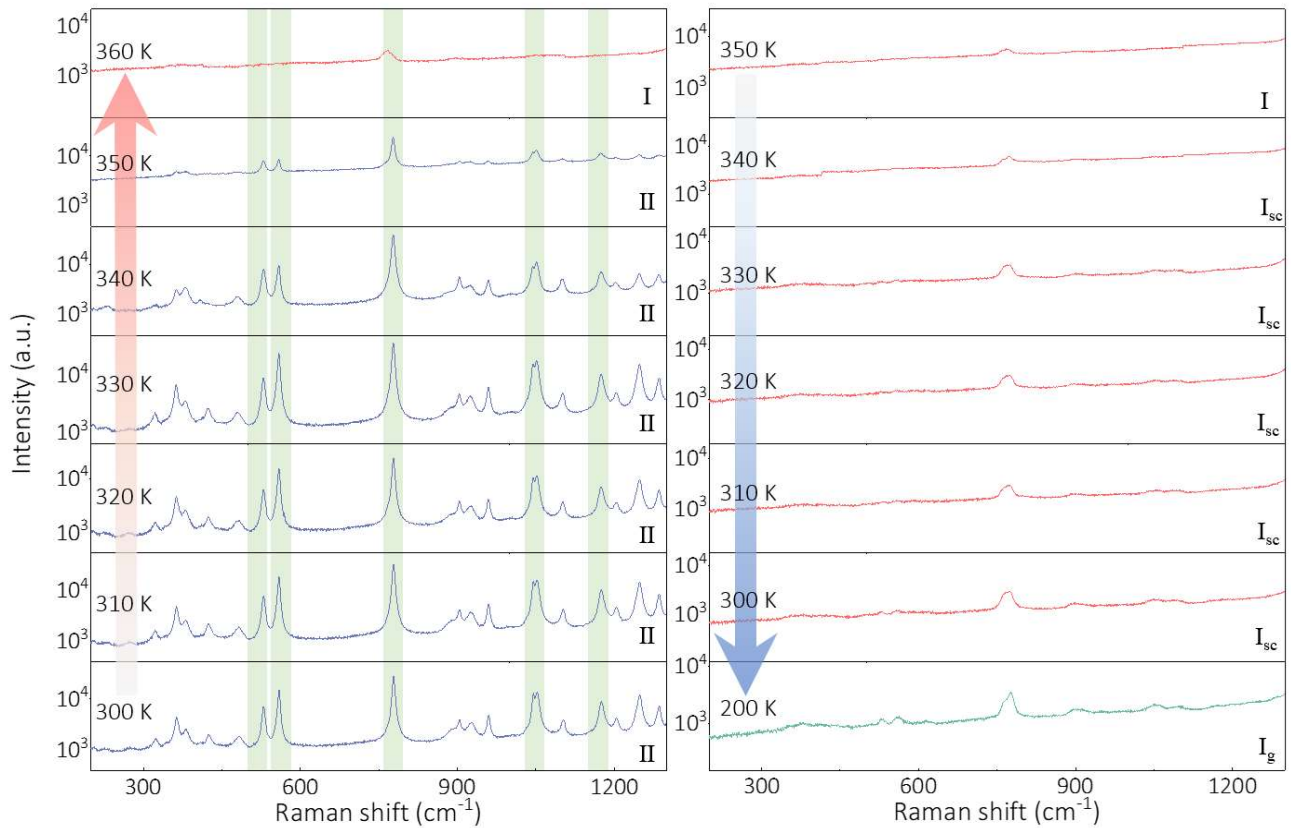

**Extended Data Fig. 2 Temperature-dependent Raman spectra.** The red (blue) arrows represent the heating (cooling) runs. In the heating run, five strong peaks (highlighted) are observed at room temperature while at 360 K, most peaks disappear and only one strong peak is found at 765 cm$^{-1}$. Thus, it is confirmed that the II to I phase transition takes place. With cooling down to 200 K, the basic profiles of the Raman spectra are the same, which manifests that the cubic lattice symmetry is retained. It is noticeable that the peak at 765 cm$^{-1}$ is slightly shifted to larger frequencies. This is a common hardening effect due to suppressed thermal fluctuations. For the same reason, some weaker peaks become visible at 200 K, for example, at ~562 cm$^{-1}$.

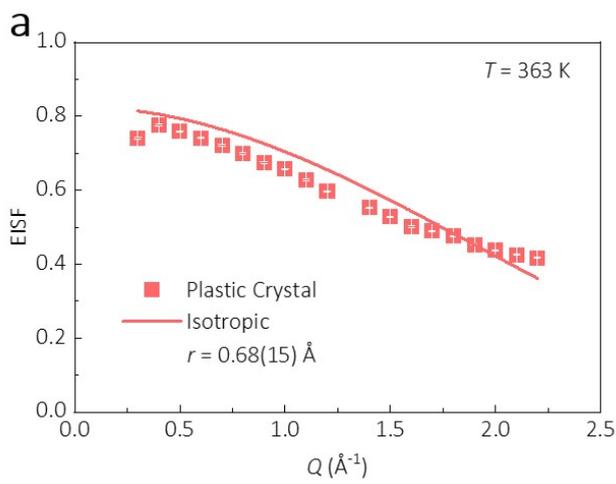 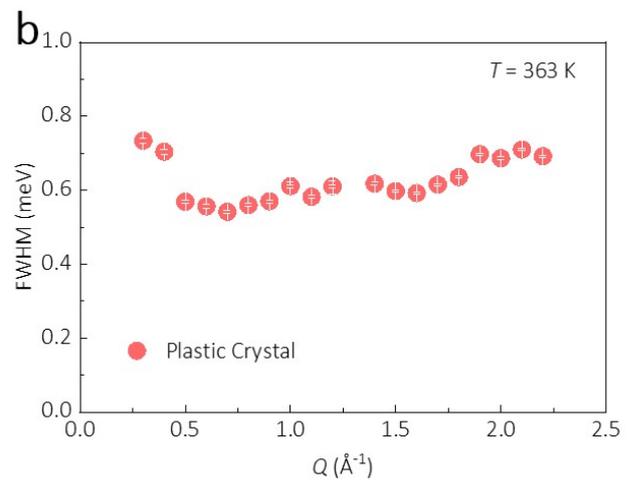

**Extended Data Fig. 3 QENS fitting results of the plastic-crystal state. a**. The experimental EISF at 363 K fitted to the isotropic reorientation model. *r*, rotational radius. **b**. The $Q$ dependence of the full width at half maximum (FWHM) of the Lorentzian component for the isotropic reorientation model.

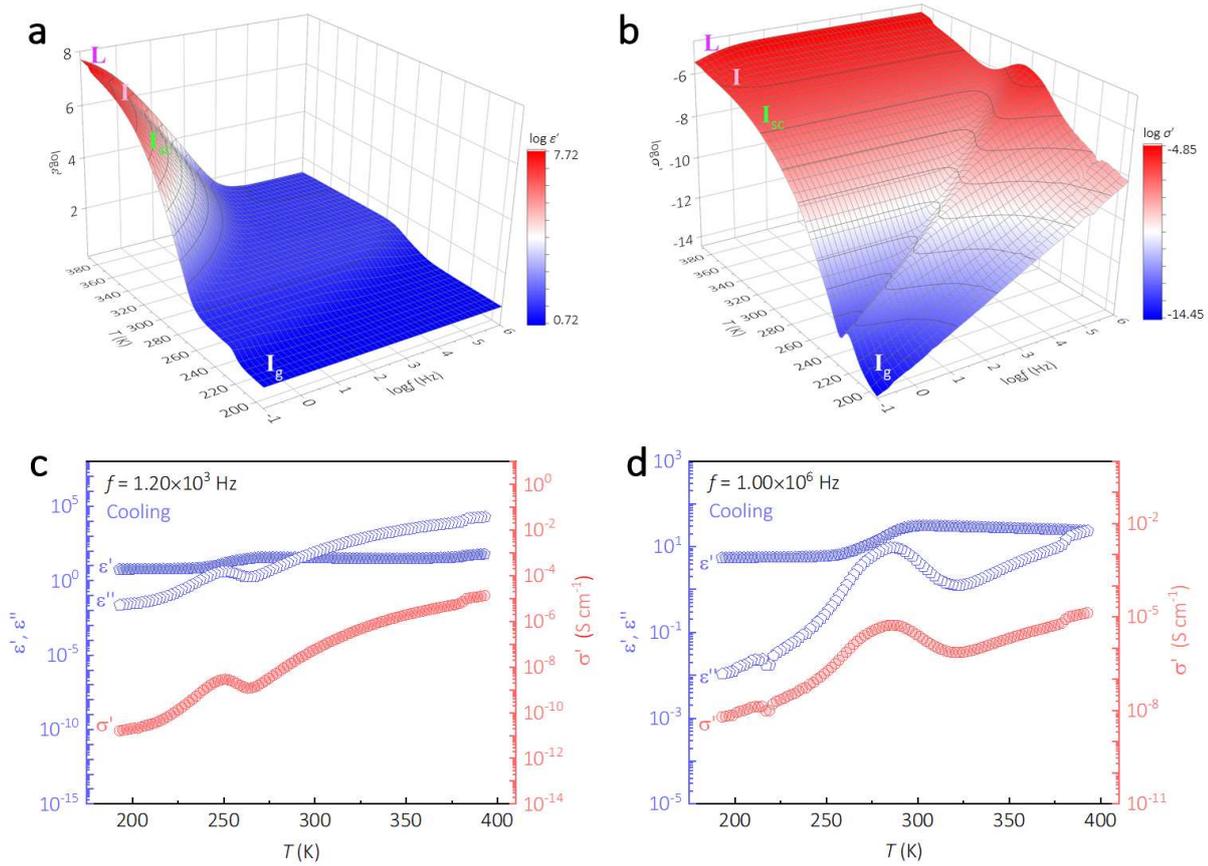

**Extended Data Fig. 4** Surface plot of the dielectric permittivity $\varepsilon'$ (**a**) and the real part of the complex electrical conductivity $\sigma'$ (**b**) upon cooling at frequencies from $10^{-1} - 10^{6}$ Hz. The temperature-dependent dielectric properties at the frequency of $1.20 \times 10^{3}$ Hz (**c**), and $1.00 \times 10^{6}$ Hz (**d**).

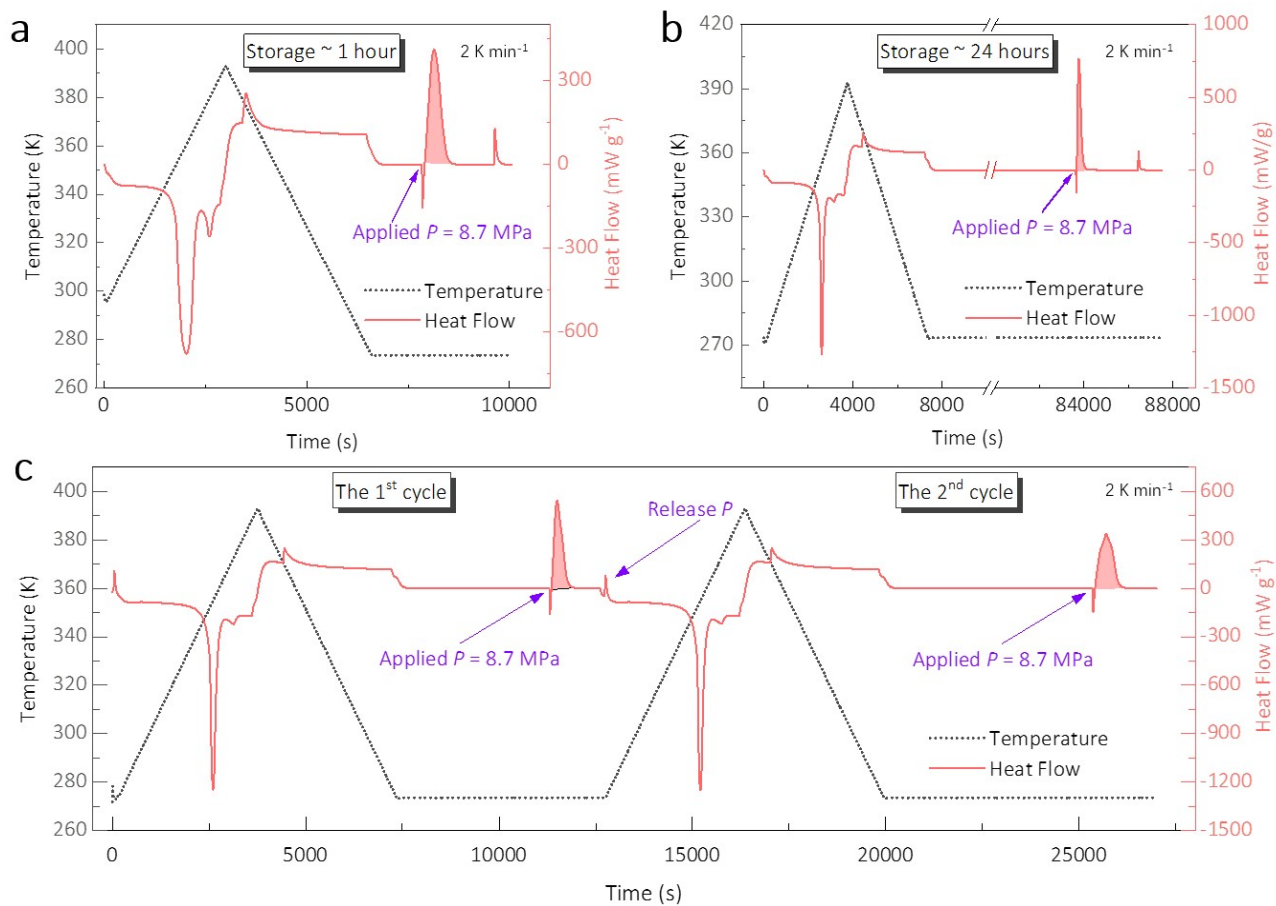

**Extended Data Fig. 5 Additional thermal charging and discharging data**. The repeated heat flow variations as a function of time for the thermal battery process obtained in AMP. The moments for applying and releasing pressures are arrowed. The thermal discharging temperatures under pressure are all at 273 K. **a**. The heat flow data with a storage time of ~ 1 hour. **b**. The heat flow data with a storage time of ~ 24 h. **c**. The thermal battery processes in one sample are repeated for two cycles continuously.

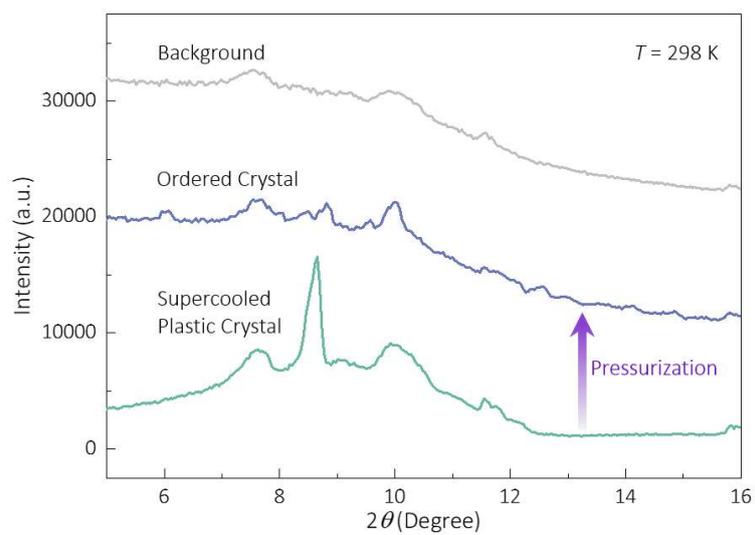

**Extended Data Fig. 6 Pressure-induced phase transition.** In-situ XRD patterns under ambient pressure and a pressure of 8 MPa, at 303 K.

**Extended Data Table 1 Comparison of the heat storage performance.**

| | Materials | Energy storage density (J g$^{-1}$) | Field (MPa) | Ref. |
|---|---|---|---|---|
| Pressure-controlled | AMP | 134 | 6.7 | This work |
| | λ-Ti$_3$O$_5$ | 52 | 60 | 31 |
| | Sc$_x$Ti$_{3-x}$O$_5$ | 16 | 1700 | 32 |
| | NH$_4$SCN | 43 | ~ 250 | 24 |
| Stress-controlled | NiTi | 35 | ~ Hundreds | 33 |
| Light-controlled | PCM+cis-Azo | 200 | -- | 34 |

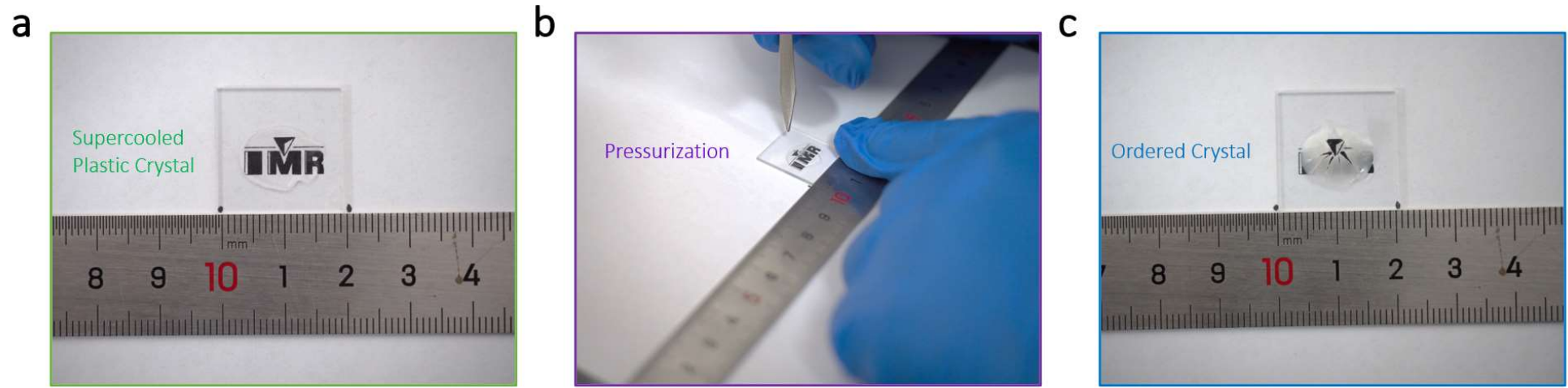

**Extended Data Fig. 7 The appearance of AMP in different states obtained by the camera. a.** The as-prepared supercooled state is transparent. **b**. The as-prepared supercooled sample is being needled. **c**. The pressure-induced ordered-crystal state is white. The cracks are formed due to accumulated strains.

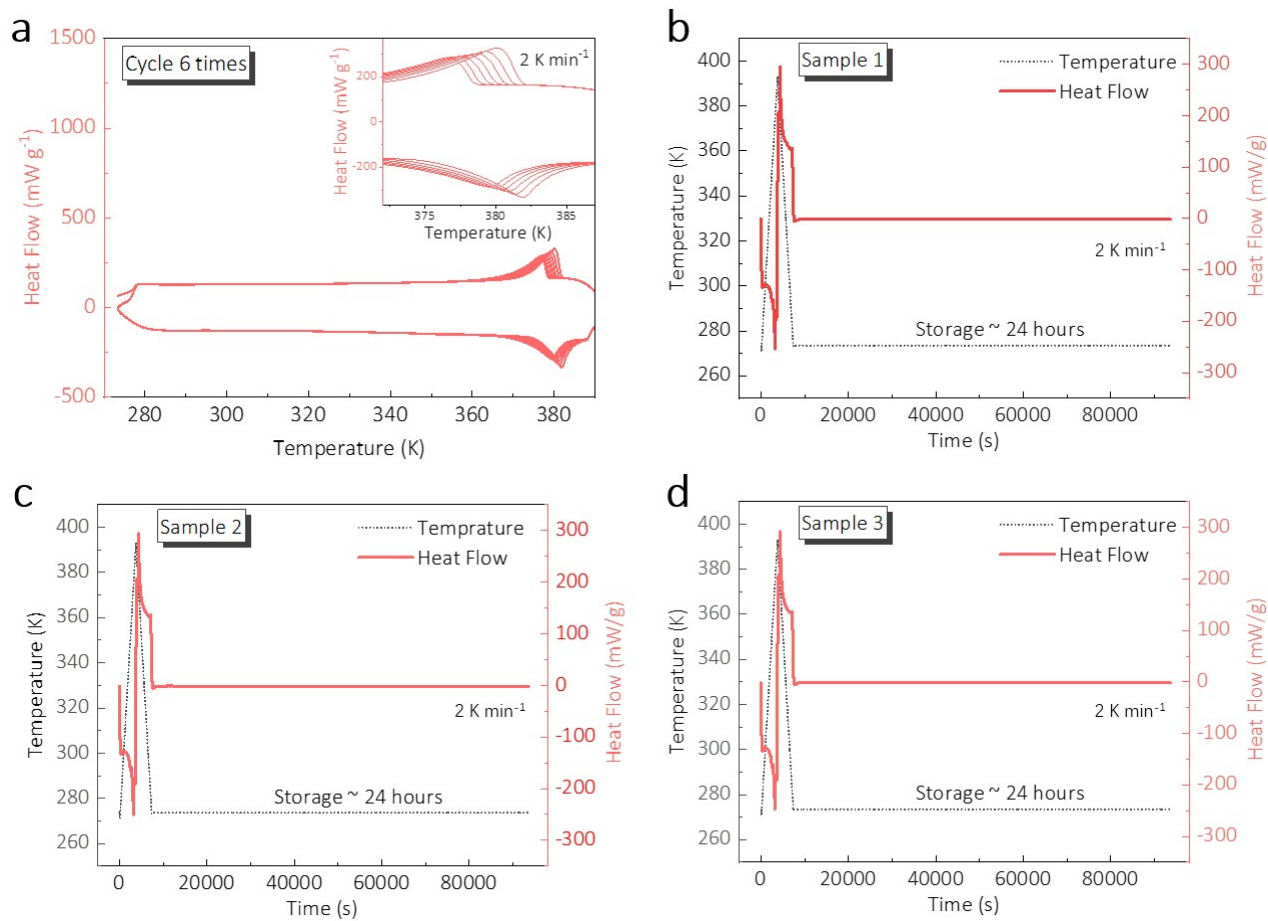

**Extended Data Fig. 8 Cycle stability and reproducibility. a.** The heat flow data of an AMP sample in the supercooled plastic crystals state repeated for six cycles between the temperature range of 273 to 393 K. The heat flow data with a storage time of ~24 h for three different samples, sample 1 (**b**), sample 2 (**c**), and sample 3 (**d**).

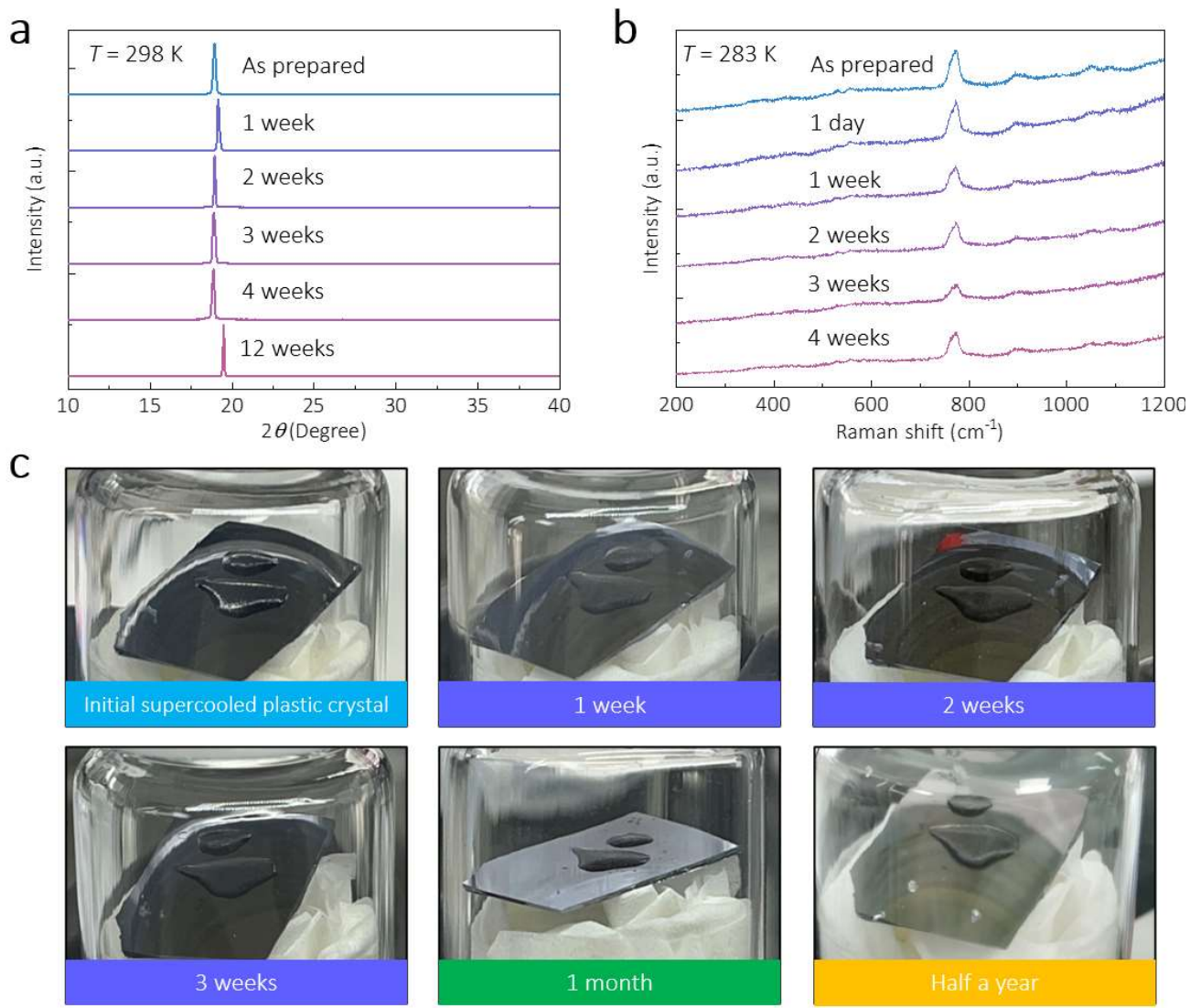

**Extended Data Fig. 9 The long-term stability of the supercooled plastic crystal state of AMP samples. a. Ex-situ XRD patterns.** Note that the shifting of the peak is because the surfaces of the samples are not flat. **b**. **In-situ Raman spectra**. **c**. Photography of AMP samples kept in a glove box.